\documentclass[]{article}  

\usepackage[]{graphicx}
\usepackage{amsmath}
\usepackage{amssymb}
\usepackage{multirow}
\usepackage{subfigure}
\usepackage{epsfig}
\usepackage{wrapfig}
\usepackage{color}
\usepackage{cite}

 \advance \textwidth by 1cm
\def\##1{{\bf #1}}
\def\=#1{\underline{\underline #1}}
\def\4#1{\underline{\underline{\underline{\underline #1}}}}
 
\def\.{\mbox{ \tiny{$^\bullet$} }}

\def\lambdao{\lambda_{\scriptscriptstyle 0}}

\def\ko{k_{\scriptscriptstyle 0}}

\def\eps{\varepsilon}

\def\.{\mbox{ \tiny{$^\bullet$} }}

\def\eps{\varepsilon}

\def\lambdao{\lambda_{\scriptscriptstyle 0}}

\def\ko{k_{\scriptscriptstyle 0}}

\def\epsa{\eps_a}
\def\epsb{\eps_b}
\def\epsc{\eps_c}

\def\chiv{\chi_v}

\def\ux{\hat{\mathbf{u}}_x}
\def\uy{\hat{\mathbf{u}}_y}
\def\uz{\hat{\mathbf{u}}_z}

\def\skipline{\vskip 5mm}

\begin{document}

\begin{center}
\textbf{Evolution of surface-plasmon-polariton and Dyakonov--Tamm  waves with the ambichirality of a partnering dielectric material}\\
\vspace{0.5cm}

{Muhammad Faryad and Akhlesh Lakhtakia }\\

\vspace{0.5cm}

{ Pennsylvania State University, Department of Engineering Science and
Mechanics, Nanoengineered Metamaterials Group (NanoMM),
University Park, PA  16802--6812, USA\\
}
\vspace{0.5cm}

 \end{center}

\pagestyle{plain}

\begin{abstract}
The planar interface of an isotropic homogeneous metal and an ambichiral dielectric material  can guide surface-plasmon-polariton waves. The planar interface of an isotropic, homogeneous dielectric material and an ambichiral dielectric material  can guide Dyakonov--Tamm waves. In either instance, we found that, as the ambichiral partnering material evolves into a finely chiral material, the solutions of the dispersion equation for surface-wave propagation evince convergence. The convergence is faster for the surface waves with larger phase speeds than for the surface waves with smaller phase speeds.

\skipline


\end{abstract}

\section{Introduction}
Anticipating the
 discovery of  cholesteric liquid crystals by about two decades \cite{Reinitzer}, Reusch proposed in 1869 that a periodically nonhomogeneous multilayered material reflects normally incident circularly polarized light of one handedness, but  not of the opposite handedness,  provided that all layers are made of the same homogeneous, uniaxial dielectric  material such that the optic axis in each layer is rotated   about the thickness direction with respect to the optic axis in the adjacent layer by a fixed angle~\cite{ReuschE}. Such a periodically nonhomogeneous dielectric material is nowadays called a \textit{Reusch pile}. 
 
 Extensive  theoretical and experimental work by Joly and colleagues ~\cite{Joly1,Joly2,Joly3,Joly4} showed that 
 circular-polarization-selective reflection of normally incident  light by a Reusch pile may occur in several spectral regimes. 
 This selective reflection of circularly polarized light of one handedness, but very little  of the other, in a given spectral regime
 is commonly called circular Bragg phenomenon~\cite{STFbook}. 

According to a classification scheme developed by Hodgkinson {\it et al.}~\cite{HodgkinsonOC04}, if the number of layers in each period $N=2$, the Reusch pile can be called an equichiral material; if $N>2$, but not very large, it can be called an ambichiral material; and if $N\rightarrow\infty$, it  is a finely chiral material. Equichiral materials do not exhibit the circular Bragg phenomenon. Ambichiral materials may exhibit the circular Bragg phenomenon in several spectral regimes, depending on the variations of their constitutive parameters with frequency. Therefore, a cholesteric liquid crystal \cite{deG} can be considered as a finely chiral Reusch pile made of uniaxial dielectric layers.

Reusch piles can also be made of biaxial dielectric material such as columnar thin films (CTFs) \cite{HWbook}.   A chiral sculptured thin film (STF) \cite{STFbook} can be considered a finely chiral Reusch pile comprising biaxial CTFs. Chiral STFs were first fabricated by Young and Kowal~\cite{Young} in 1959 and were rediscovered in the 1990s~\cite{Robbie1995}. They have been extensively studied since then for optical applications exploiting the circular Bragg phenomenon~\cite{STFbook}.

The effect of the number of layers $N$ on the circular Bragg phenomenon has been studied~\cite{Abdulhalim2008}.   Both $N$ and
 the total number of periods have to be substantially large for the circular Bragg phenomenon to fully develop~\cite{Abdulhalim2008}. What is the effect of $N$ on the surface-wave propagation guided by the interface of a homogeneous isotropic material and an ambichiral dielectric material? The results reported in this Letter are due to the investigations conducted to answer that question. 
 
 The planar interface of an isotropic homogeneous metal and an ambichiral dielectric material  can guide surface-plasmon-polariton waves. The planar interface of an isotropic, homogeneous dielectric material and an ambichiral dielectric material  can guide Dyakonov--Tamm waves. For surface waves of both types, we examined the evolution of the solution(s) of the dispersion equation with $N$.

\section{Theoretical Preliminaries}
\begin{figure}[!ht]
\begin{center}
\includegraphics[width=3.0in]{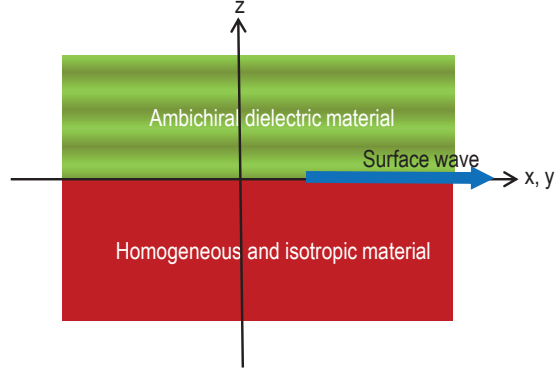}
\caption{Schematic of the canonical boundary-value problem.}
\label{canonical}
\end{center}
\end{figure}

For this purpose, we considered the   canonical boundary-value problem of surface-wave propagation,  shown schematically in Fig.~\ref{canonical}. The half space $z<0$ is occupied by isotropic and homogeneous material with relative permittivity $\eps_{s}$.
The half space $z>0$ is occupied by an ambichiral dielectric material comprising homogeneous layers each of thickness $D$, the  $\ell$th layer occupying the region $(\ell-1)D < z < {\ell}D$, $\ell\in[1,\infty)$. The relative permittivity dyadic is given as
\begin{multline}
\=\epsilon\left(z,\omega\right)= \=S_z\left(h\xi_\ell+h\gamma\right)\.\=S_y(\chi)\.
 \=\epsilon _{ref}^\circ\left(\omega\right) \.\=S_y^{-1}(\chi)
 \. \=S_{z}^{-1}\left(h\xi_\ell+h\gamma\right)\,,\\
 (\ell-1)D < z < {\ell}D\,,~
\ell\in[1,\infty)\,,
\label{perm_reusch}
\end{multline}
where
the reference permittivity dyadic  
\begin{equation}
\=\eps_{ref}^\circ\left(\omega\right)= 
\uz\uz \,\epsa(\omega) +\ux\ux\,\epsb(\omega)+ \uy\uy \,\epsc(\omega)
\label{epsref_CTF-Reusch}
\end{equation}
contains the eigenvalues $\eps_{a,b,c}(\omega)$ of $\=\epsilon\left(z,\omega\right)$,
 the dyadic
\begin{equation}
\=S_y(\chi)= (\ux\ux+\uz\uz)\cos\chi+(\uz\ux-\ux\uz)\sin\chi+\uy\uy
\end{equation}
depends on the tilt angle $\chi\in[0^\circ,90^\circ]$ woth respect to the $xy$ plane,
 the dyadic
\begin{equation}
\=S_{z}\left(\xi\right)=(\ux\ux+\uy\uy) \cos\left(\xi\right) +
(\uy\ux - \ux\uy) \sin\left(\xi\right) +\uz\uz
\label{Szpile}
\end{equation}
represents a rotation about the $z$ axis by an angle $\xi$,
$\xi_\ell=(\ell-1)\pi/N$ with $N\geq1$ being the
number of  layers in each period $2\Omega=ND$, 
right-handed rotation is represented by $h=1$ and left-handed rotation  by $h=-1$,  and
$\gamma$ is an angular offset with respect to the $x$ axis.

Without any loss of generality, let us suppose that the surface wave propagates parallel to the $x$ axis guided by the interface plane $z=0$.  The associated electromagnetic fields depend on $x$ as
 $\exp(iqx)$, have no dependence on $y$, and their dependences on $z$ in both half spaces  indicate decay  as
 $\vert{z}\vert\to\infty$.
 The wavenumber $q$ is complex valued in general.
The complete formulation
of the canonical boundary-value problem to obtain a dispersion equation for $q$ 
being available elsewhere~\cite{LP2007,PLprsa,PMLbook},  we proceed directly to the presentation of numerical results.

\section{Numerical results and discussion}\label{nrd}
The dispersion equation was solved using the Newton--Raphson method~\cite{Jaluria},
with the free-space wavelength $\lambdao$ fixed at $633$~nm.  
For all numerical results presented here, the ambichiral dielectric material was taken to 
comprise CTFs made by evaporating patinal titanium oxide \cite{HWH} by directing a collimated evaporant flux
in a low-pressure chamber at a fixed angle  $\chiv\in(0^\circ,90^\circ]$
with respect to the planar substrate. For the chosen CTF,
\begin{equation}
\left.\begin{array}{l}
\epsa=\left[1.0443+2.7394 \left(\frac{2\chiv}{\pi}\right)-1.3697 \left(\frac{2\chiv}{\pi}\right)^2\right]^2\\[5pt]
\epsb=\left[1.6765+1.5649 \left(\frac{2\chiv}{\pi}\right)-0.7825 \left(\frac{2\chiv}{\pi}\right)\right]^2\\[5pt]
\epsc=\left[1.3586+2.1109 \left(\frac{2\chiv}{\pi}\right)-1.0554 \left(\frac{2\chiv}{\pi}\right)^2\right]^2\\[5pt]
\chi=\tan^{-1}\left(2.8818\tan\chiv\right)
\end{array}\right\}\,
\label{cstf_per}
\end{equation}
according to  Hodgkinson and co-workers \cite{HWH}. We fixed $\Omega=200$~nm, while varying
$N\in[1,15]$ (so that $D=2\Omega/N$ was simultaneously varied) and $\gamma\in\left\{0^\circ,45^\circ,90^\circ\right\}$.

\begin{figure}[!ht]
\begin{center}
\includegraphics[width=5.50in]{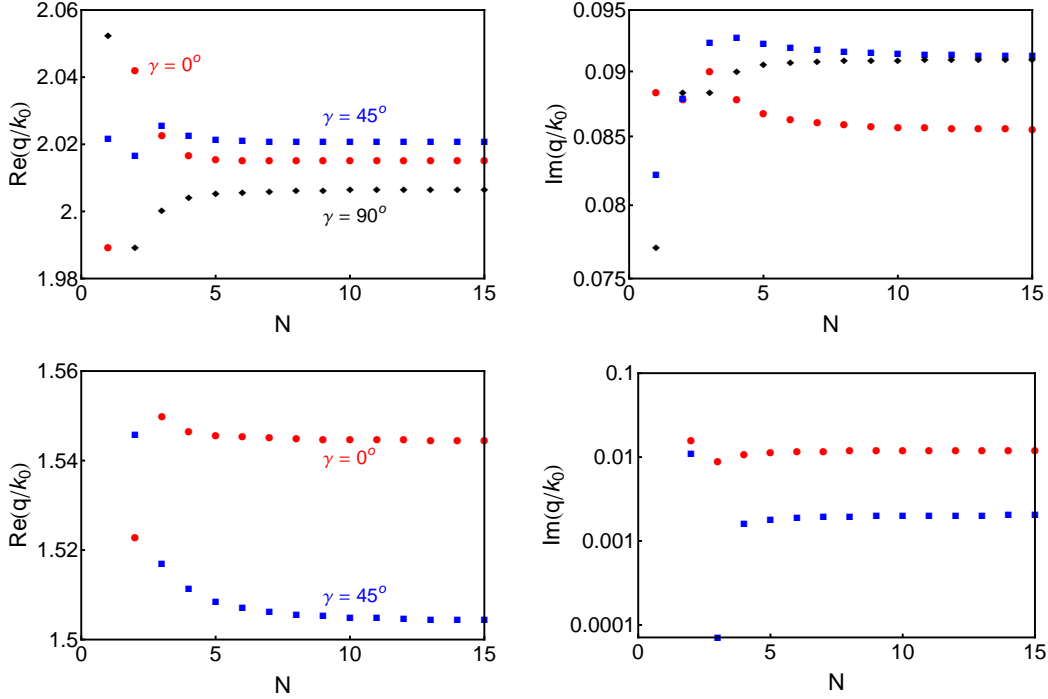}
\caption{Real and imaginary parts of the relative wavenumbers $q/\ko$ calculated as solutions
of the surface-wave dispersion equation
as a function of $N\in[1,15]$ and $\gamma\in\left\{0^\circ,45^\circ,90^\circ\right\}$
for SPP waves guided by the interface
of aluminum ($\eps_s=-14.65+i5.85$) and an ambichiral dielectric
material characterized by Eqs.~(\ref{cstf_per}) with $\chiv=20^\circ$. For these
calculations, we fixed  $\lambdao=633$~nm and $\Omega=200$~nm. Top: First solution. Bottom: Second solution.}
\label{spp1}
\end{center}
\end{figure}

\subsection{SPP waves}
Let the  isotropic  homogeneous  partnering material be thin-film aluminum  with 
$\eps_s=-14.65+i5.85$ \cite{PartIII} at $\lambdao=633$~nm. All solutions of the surface-wave dispersion
equation depicted  as functions of $N$  in Fig.~\ref{spp1} represent 
surface-plasmon-polariton (SPP) waves~\cite{PLprsa,PMLbook}. These solutions are complex valued, as necessitated
by $\eps_s$ being complex valued.

Only one solution of the dispersion equation exists when the anisotropic partner is homogeneous ($N=1$) \cite{PLoc}.
However, two solutions exist when that partner is periodically nonhomogeneous ($N>1$) when $\gamma=0^\circ$ and $45^\circ$. For $\gamma=90^\circ$, only one solution was found.

The most notable feature of the solutions presented in Fig.~\ref{spp1}   is their evolution  as $N$ increases. Specifically, the solutions $q$ in Fig.~\ref{spp1}(top) converge to within $\pm0.1\%$  of the corresponding solutions for the
metal/chiral-STF interface \cite{PLprsa} when $N\geq18$, and  the solutions  in Fig.~\ref{spp1}(bottom) converge to within $\pm0.1\%$ when $N\geq55$. Also, 
${\rm Re}(q)$ converges faster  than ${\rm Im}(q)$.

\subsection{Dyakonov--Tamm waves}
\begin{figure}[!ht]
\begin{center}
\includegraphics[width=3.0in]{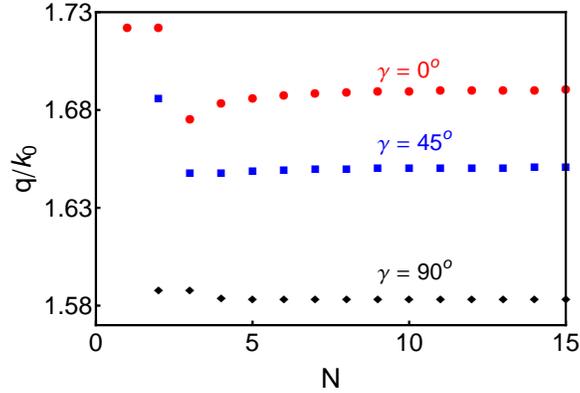}
\caption{The  wavenumbers $q/\ko$ calculated as solutions
of the surface-wave dispersion equation
as a function of $N\in[1,15]$ and $\gamma\in\left\{0^\circ,45^\circ,90^\circ\right\}$
for Dyakonov--Tamm waves guided by the interface
of magnesium fluoride ($\eps_s=1.896$) and an ambichiral dielectric
material characterized by Eqs.~(\ref{cstf_per}) with $\chiv=20^\circ$. For these
calculations, we fixed  $\lambdao=633$~nm and $\Omega=200$~nm. }
\label{dt}
\end{center}
\end{figure}

Next, let the   isotropic  homogeneous  partnering material be  magnesium fluoride---a dielectric material  with 
$\eps_s=1.896$ at $\lambdao=633$~nm. 
All solutions of the surface-wave dispersion
equation depicted  as functions of $N$  in Fig.~\ref{dt} represent 
Dyakonov--Tamm waves~\cite{LP2007,PMLbook}.
All the solutions $q$ are real valued, because  dissipation is negligibly small in both  partnering materials. 

For $N=1$, the anistropic partner is homogeneous and the solution $q$ when $\gamma=0^\circ$ represents a Dyakonov wave~\cite{Dyakonov,PMLbook}. For $N>1$, the solutions $q$ represent  Dyakonov--Tamm waves. Figure \ref{dt} indicates the typical difference between Dyakonov and Dyakonov--Tamm waves: The range of $\gamma$ is much larger for surface waves of the latter type than for the  surface waves of the former type.

Just like the solutions presented in Fig.~\ref{spp1}   for SPP waves, those presented in Fig.~\ref{dt} for Dyakonov--Tamm waves also evolve as $N$ increases. Specifically, the solutions for Dyakonov--Tamm waves in Fig.~\ref{dt} converge to within $0.1\%$ of the corresponding solution for the
isotropic-dielectric/chiral-STF interface \cite{LP2007}  when $N\geq29$.

\section{Concluding Remarks}\label{conc}
The canonical boundary-value problem of surface-wave propagation guided by the planar of  an isotropic  homogeneous material and an ambichiral material was set up and solved. Both Dyakonov--Tamm and SPP waves were investigated. As the number 
$N$ of CTFs per period in the ambichiral partnering material increased, the solutions of the surface-wave dispersion equation were found to converge to those for the ambichiral material replaced by the corresponding finely chiral material. The convergence was faster when the isotropic  homogeneous partner was a dielectric material than when it was a metal. For SPP waves, the real part of the surface wavenumber $q$ converge faster than the imaginary part. The solutions converge faster for the surface waves with smaller phase speeds than for the surface waves with larger phase speeds, irrespective of the fact that isotropic homogeneous partner is a metal or a dielectric material.

\vspace{0.5cm}
MF and AL are grateful for partial support from Grant No.  DMR-1125591
of the US National Science Foundation.
AL is also grateful to  the Charles Godfrey Binder Endowment
at the Pennsylvania State University for partial support of this work.

\  \end{document}